\documentclass[epjST]{svjour}
\usepackage{graphics}
\usepackage{hyperref}
\usepackage{amsmath,amssymb}

\newcommand{\vect}[1]{\boldsymbol{#1}}

\begin{document}
\title{Applications of the Nambu--Jona-Lasinio model to the partonic structure of the pion\thanks{Dedicated 
to David Blaschke on the occasion of his 60th birthday. Presented by WB at 
The 40th Max Born Symposium -- Three Days on Strong Correlations in Dense Matter, 9-12 October 2019, Wroc\l{}aw, 
Poland.}\fnmsep\thanks{Supported by the Polish National Science Centre (NCN)
Grant 2018/31/B/ST2/01022, the Spanish Ministerio de Economia y
Competitividad and European FEDER funds Grant FIS2017-85053-C2-1-P,
and Junta de Andaluc\'{\i}a Grant FQM-225.}}
\author{Wojciech Broniowski\inst{1}\fnmsep\inst{2}\fnmsep\thanks{\email{Wojciech.Broniowski@ifj.edu.pl}}
 \and Enrique Ruiz Arriola\inst{3} \fnmsep\thanks{\email{earriola@ugr.es}}}
\institute{Institute of Physics, Jan Kochanowski University, 25-406 Kielce, Poland \and 
                Institute of Nuclear Physics PAN, 31-342 Cracow, Poland \and 
                Departamento de F\'isica At\'omica, Molecular y Nuclear and Instituto Carlos I de F\'{\i}sica Te\'orica y Computacional,
                   Universidad de Granada, E-18071 Granada, Spain}
\abstract{
We present a brief review of results of chiral quark models for soft matrix elements in the pion state, 
appearing in high-energy processes as well 
as accessible in present and future lattice studies. A particular 
attention is paid to the recently explored double parton distribution functions of the pion. 
}
\maketitle

\section{Introduction}
\label{sec:intro}

David has been successfully using quark models all over his career,
including applications to dense matter where the otherwise usually
confined quarks may be de-confined and give rise to stars with a quark
core~\cite{Alford:2006vz,Shahrbaf:2019vtf}. In this talk we focus on
other aspects of chiral quark models, such as the Nambu--Jona-Lasinio
(NJL) model, applied to the domain where they were originally designed
for, namely the soft limit in the vacuum. There the quarks remain
confined, but the chiral symmetry is broken, leading to rich dynamical
predictions.  It is not so commonly known that in this case the model
explains numerous features of the pion (in general, the Goldstone
bosons), both in low- and high-energy processes, whenever the soft-hard
factorization holds (see~\cite{RuizArriola:2002wr} for a detailed
review).  These results, amended with the necessary QCD evolution
which generates radiatively the gluon degrees of freedom, compares
very favorably to the available experimental and lattice data.  We 
briefly review some of these results and then pass to a recent topic
of double parton distribution functions (dPDF) in the
pion~\cite{BW-ERA-LC2019,Courtoy:2019cxq,Broniowski:2019rmu} evaluated
in chiral quark models followed with the DGLAP evolution. We discuss
the issue of partonic correlations and the proposed measures based on the
Mellin moments of dPDFs~\cite{Broniowski:2019rmu}.

\section{Basic formalism}
\label{sec:sPDF}

The field-theoretic definition of the single parton distribution function  (sPDF)  involves a
diagonal matrix element of bilinear operators 
in a hadronic state (see \cite{Diehl:2010dr} and references therein), namely
\begin{eqnarray}
  && D_{j}(x) =   \int \frac{d z^-}{2\pi}\,
    e^{i x^{} z^- p_{}^+} \langle p |\,
   { \mathcal{O}_{j}(0,z)} \,| p \rangle
    \bigl|_{z^+ = 0\,, \vect{z}^{} = \vect{0}}.
\end{eqnarray}
Here $p$ is the momentum of the hadron, $x$ is the Bjorken variable interpreted as the fraction of the light-cone momentum of the 
hadron carried out  by the struck parton, and $ \mathcal{O}_{j}(0,z)$ is a bilocal operator which for the quarks and anti-quarks considered here 
and in the applied light-cone gauge takes the form 
\begin{eqnarray}
&& \hspace{-2mm} \mathcal{O}_{q}(y, z) = \tfrac{1}{2}\, \bar{q} ( y- \tfrac{z}{2}) \gamma^+  q ( y+ \tfrac{z}{2} ), \nonumber \\
&& \hspace{-2mm} \mathcal{O}_{\bar{q}}(y, z) =-\tfrac{1}{2}\, \bar{q} ( y + \tfrac{z}{2}) \gamma^+  q ( y - \tfrac{z}{2}).
\end{eqnarray} 
The light-cone coordinates are introduced as $v^\pm = (v^0 \pm v^3) /\sqrt{2}$, whereas the boldface indicates the transverse components, 
$\vect{v}=(v^1, v^2)$. The quark-pion coupling is point-like, as follows from the NJL model.

\begin{figure}
\begin{center}
\resizebox{0.4\columnwidth}{!}{\includegraphics{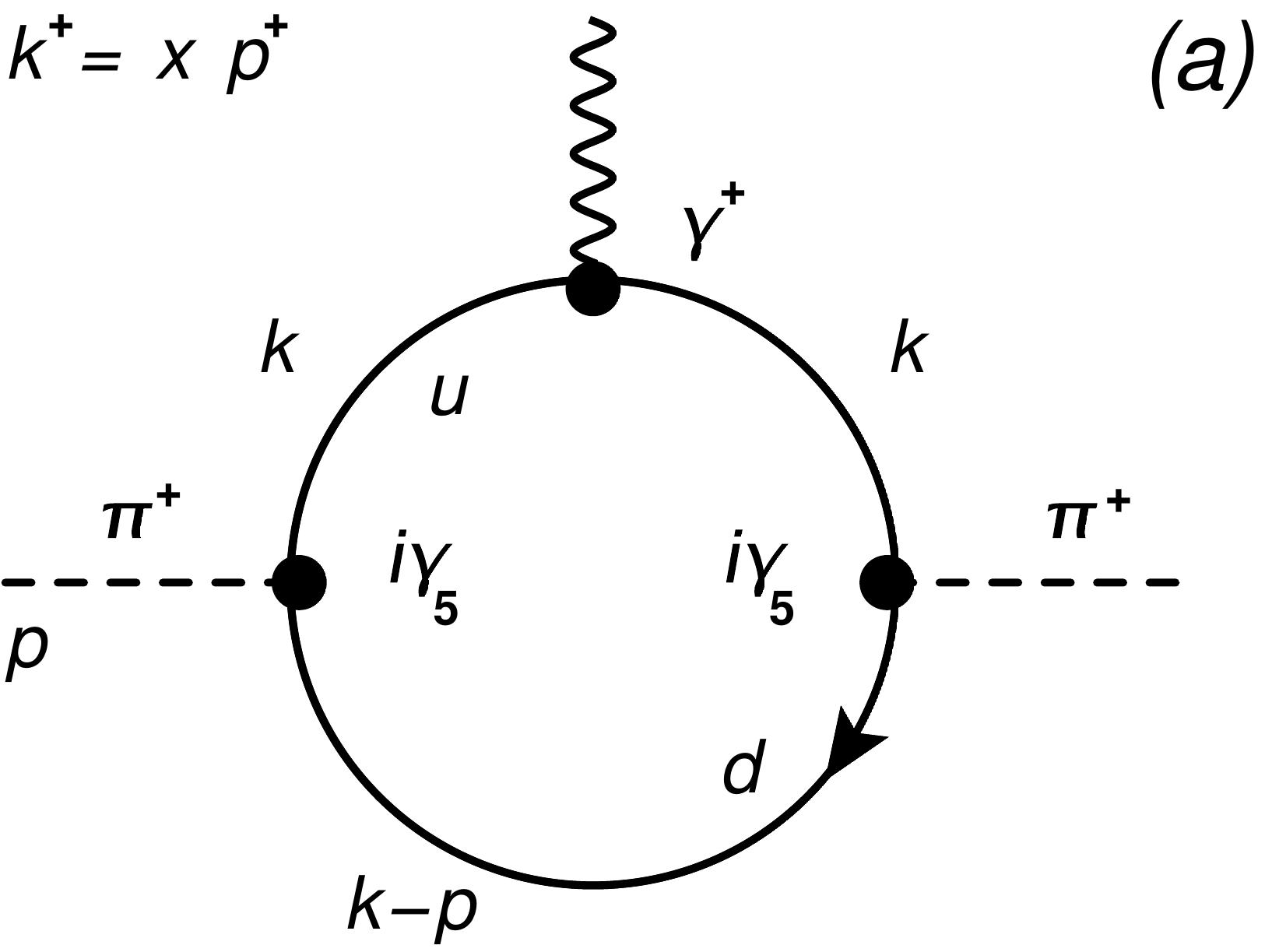} } \hspace{1.2cm}
\resizebox{0.4\columnwidth}{!}{\includegraphics{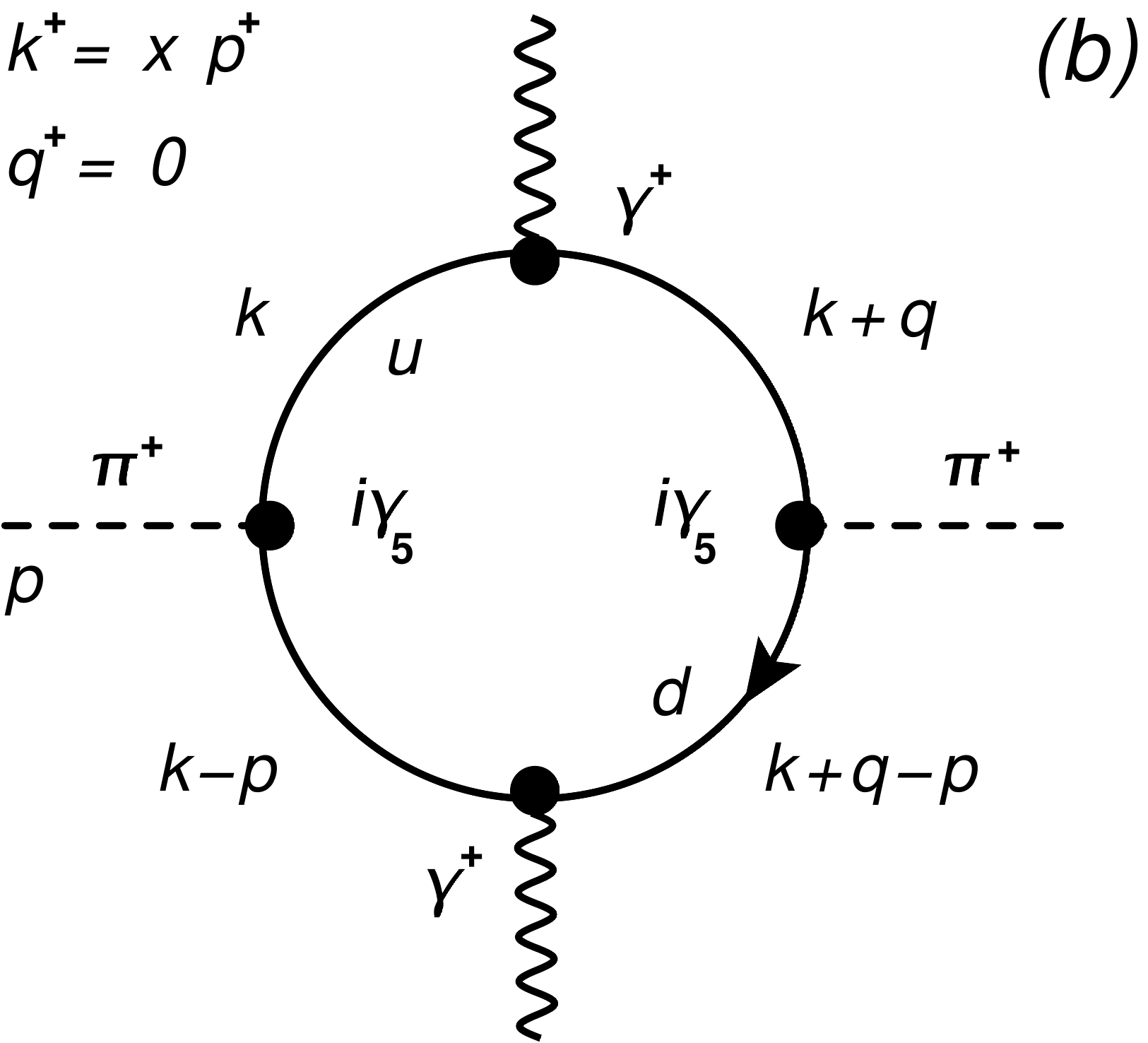} }
\end{center}
\caption{Diagrams to evaluate the single (a) and double (b) valence quark distributions of $\pi^+$ at the leading-$N_c$ order in the NJL model.}
In diagram (b) integration over $q^-$ is carried out.
\label{fig:diag}     
\end{figure}

For the double parton distribution functions (dPDF) one has, analogously, a matrix element involving two bilocal currents~\cite{Diehl:2010dr},
\begin{eqnarray}
&&  \hspace{-2mm} D_{j_{1} j_{2}}(x_1,x_2,\vect{b})
 =  2 p^+ \int d y^-\,
        \frac{d z^-_1}{2\pi}\, \frac{d z^-_2}{2\pi}\;
          e^{i ( x_1^{} z_1^- + x_2^{} z_2^-) p_{}^+} \nonumber \\
 && ~~~~ \times
    \langle p |\, {\mathcal{O}_{j_1}(y,z_1)\, \mathcal{O}_{j_2}(0,z_2)}
    \,| p \rangle
    \bigl|_{z_1^+ = z_2^+ = y_{\phantom{1}}^+ = 0\,,
    \vect{z}_1^{} = \vect{z}_2^{} = \vect{0}}, \label{eq:defsd}
\end{eqnarray}
where indices $1,2$ refer to the two partons. Note that there is an extra argument of this object, 
namely, the relative transverse distance between the partons, $\vect{b}$.

The simple meaning of the above definitions in the momentum space is illustrated in Fig.~\ref{fig:diag}, where for definiteness 
we take the case of the charged pion and use the large-$N_c$ limit, which amounts to evaluating the one quark loop. 
We note that for dPDF of Fig.~\ref{fig:diag}(b) there is a momentum flow between the two probing operators. Integration over 
$q^-$ imposes the constraint $y^+=0$ from Eq.~(\ref{eq:defsd}), whereas the transverse component $\vec{q}$ is the Fourier-conjugate
variable corresponding to $\vec{b}$. The evaluation leads to simple results presented in the sections below.

As discussed extensively in~\cite{RuizArriola:2002wr}, the evaluation according to the diagrams of Fig.~\ref{fig:diag} 
corresponds to the {\em quark model scale} $\mu_0$, where no gluons are present and 
the constituent quark and anti-quark  saturate the momentum sum rule, 
$\langle x \rangle_{\mu_0}=1$. Fits to phenomenological  sPDF provide the value $\mu_0 \sim 320~{\rm MeV}$ (see,
e.g.,~\cite{Broniowski:2007si}). The matching condition between
the model and QCD for an observable $A$  is imposed by the condition
\begin{equation}
\left . A(x,\mu_0) \right |_{\rm model} = \left . A(x,\mu_0) \right |_{\rm QCD}.
\end{equation}
Then, the QCD evolution is carried out to higher scales $\mu$, where comparison to the experimental data of lattice simulations is possible. 
This process, here performed at the leading order, generates radiatively the gluons. 
The QCD evolution is a crucial ingredient of the approach.

\section{Single parton distributions of the pion}
\label{sec:sPDF}

\begin{figure}
\begin{center}
\resizebox{0.47\columnwidth}{!}{\includegraphics{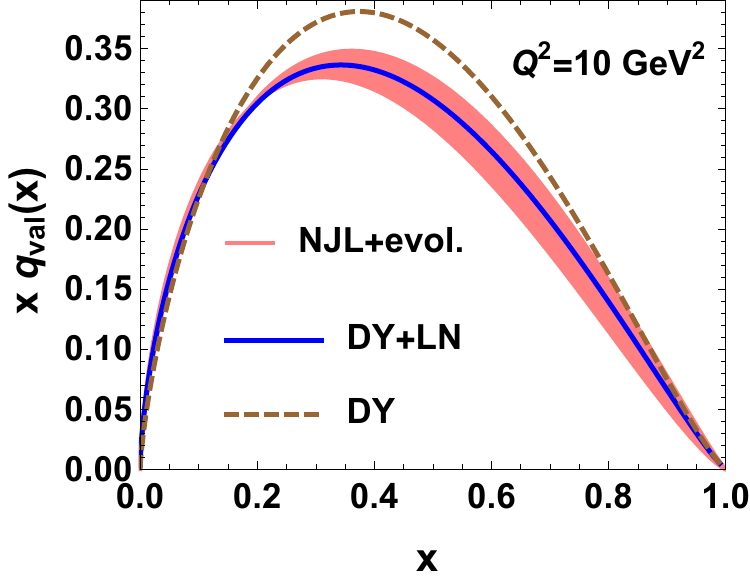} }
\end{center}
\caption{Comparison of the NJL mode followed with LO DGLAP evolution (band) to
  the extraction of the valence pion sPDF made
  in~\cite{Barry:2018ort}. The dashed line uses the $\pi A$ Drell-Yan
  data, and the solid line combines them with the HERA leading-neutron
  electro-production data.}
\label{fig:JLab}     
\end{figure}

Let us first recall the results for sPDF of the pion, first obtained
by Davidson and one of us (ERA)~\cite{Davidson:1994uv} by analyzing
the forward Compton scattering amplitude in the Bjorken
limit.~\footnote{One technical aspect of high relevance in the
  calculation is the choice of the NJL regularization, which needs
  to preserve gauge invariance and chiral symmetry, as well as their corresponding Ward-Takahashi identities.} There, at the
quark model scale and in the strict chiral limit of $m_\pi=0$, 
one gets (notation for $\pi^+$, other pion states are obtained via the isospin symmetry)
\begin{eqnarray}
q_{\rm val}(x) \equiv D_u(x)=D_{\bar{d}}(x)= 1 \times \theta(x)\theta(1-x). \label{eq:sp}
\end{eqnarray}
We note the proper support, normalization, and a uniform distribution in the ${}^+$ momentum fraction $x$. 

When evolved to higher scales, the results agree very well with the
phenomenological extractions of $q_{\rm val}(x)$ from the data.  The
comparison to the Fermilab E615 Drell-Yan data can be found, e.g.,
in~\cite{Broniowski:2007si}), so here in Fig.~\ref{fig:JLab} we
confront the model to a recent phenomenological extraction
of~\cite{Barry:2018ort}, corresponding to the scale
$\mu^2=Q^2=10$~GeV${}^2$. We note a very good agreement of the model
(the band reflects the uncertainty in the initial scale,
$\mu_0=313^{+20}_{-10}$~MeV~\cite{Broniowski:2007si}) to the analysis
of~\cite{Barry:2018ort}. In particular, we note an excellent agreement with the analysis
combining the $\pi A$ Drell-Yan data and the HERA leading-neutron
electro-production data. A very recent analysis~\cite{Novikov:2020snp} confirms
this agreement when we take $\mu_0=374$~GeV to get $0.55$ momentum fraction
of the valence quarks at $Q^2 = 5~{\rm GeV}^2$. 

Further, we wish to comment of the behavior of $q_{\rm val}(x)$ at $x \to 1$, which recently has been a subject of heated discussion. 
With the applied DGLAP evolution from $Q_0=313$~MeV, this behavior is given by~\cite{Broniowski:2007si}
\begin{eqnarray}
q_{\rm val}(x) \sim (1-x)^{4C_f/\beta_0 \log[\alpha(Q_0)/\alpha(Q)]}=(1-x)^{1.24}.
\end{eqnarray}
The power clearly evolves with the scale $Q$.
We stress a very good agreement of this behavior in our model with  the phenomenological extraction, cf.~Fig.~\ref{fig:JLab}.

Admittedly, our analysis as well as the quoted experimental extraction~\cite{Barry:2018ort} do not account for the soft-gluon resummation effects, which flatten the 
curve near the thereshold, $x\to 1$~\cite{Aicher:2010cb}. In comparisons, such effects should be included both at the theoretical and the experimental side.

There are numerous other results for the pion that can be computed with this scheme (chiral quark model followed with QCD evolution), all of them 
giving a fair comparison, whenever available, with the experimental or lattice data. Notable examples are
the parton distribution amplitude (PDA)~\cite{RuizArriola:2002bp}, the generalized parton distribution 
functions (GPD)~\cite{Broniowski:2007si}, or the quasi parton distributions~\cite{Broniowski:2017wbr,Broniowski:2017gfp}.

\section{Double parton distributions of the pion}
\label{sec:dPDF}

\begin{figure}
\begin{center}
\resizebox{0.49\columnwidth}{!}{\includegraphics{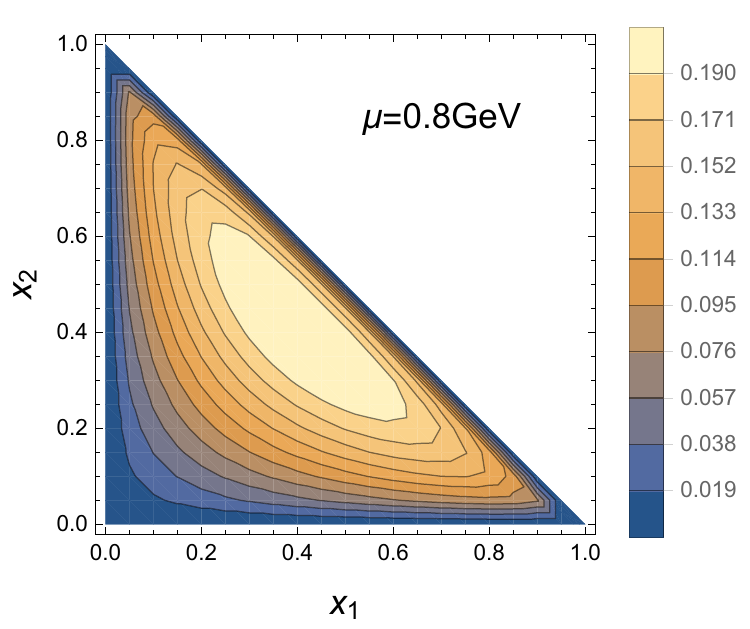}} \resizebox{0.49\columnwidth}{!}{\includegraphics{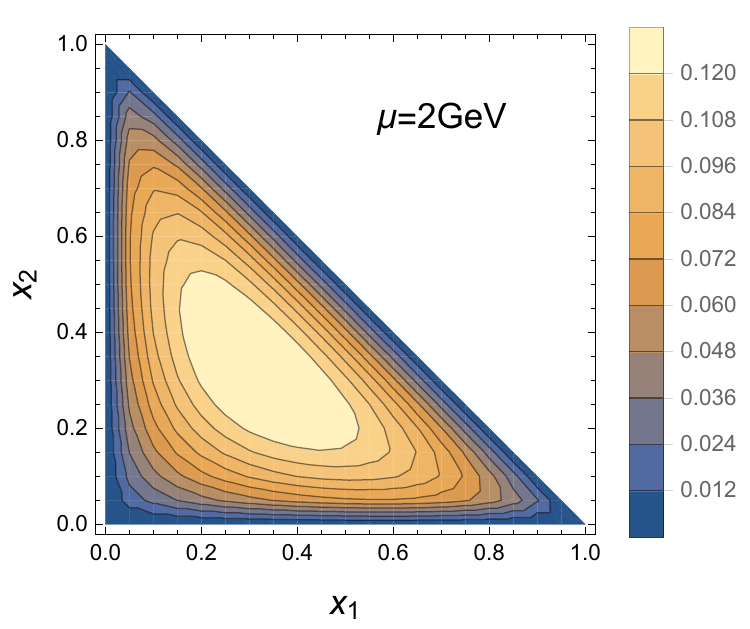}} \\
\resizebox{0.49\columnwidth}{!}{\includegraphics{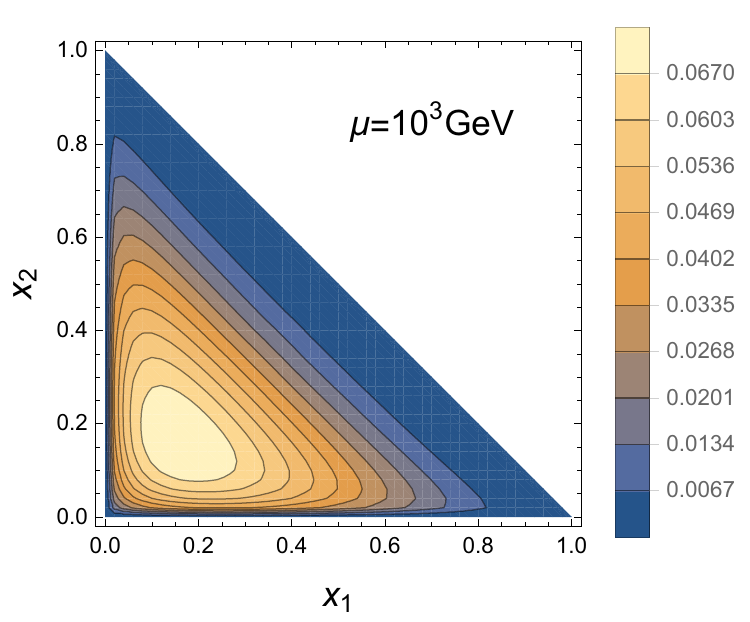}} \resizebox{0.49\columnwidth}{!}{\includegraphics{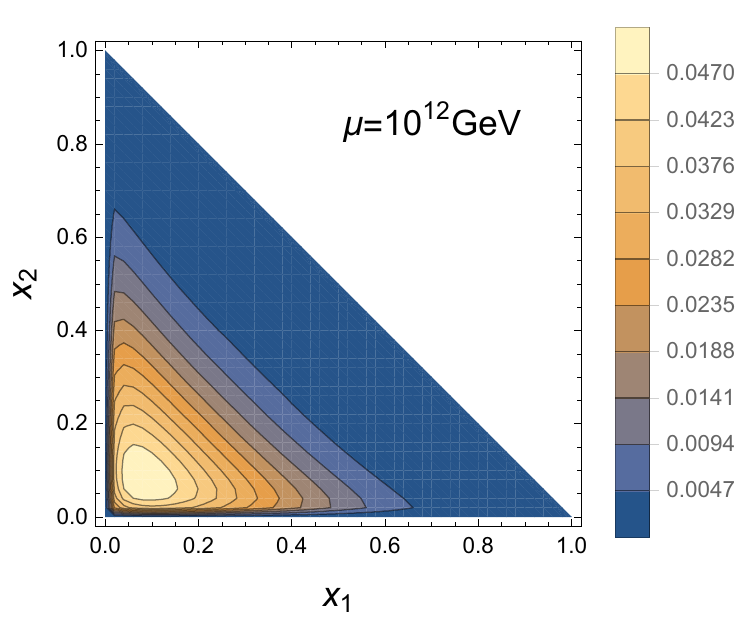}}
\end{center}
\caption{Valence dPDF of the pion, $x_1 x_2 D_{\rm u \bar{d}}(x_1,x_2)$, evolved with the DGLAP equations 
to subsequent scales $\mu$ indicated in the panels.  The initial condition at the quark model scale $\mu_0$ 
is the singular distribution of Eq.~(\ref{eq:dp})  (taken from~\cite{Broniowski:2019rmu}).  \label{fig:dev}}
\end{figure}

Now we pass to our main topic of this talk, namely, the predictions for the valence dPDF of the pion. The evaluation of the diagram from
Fig.~\ref{fig:diag}(b) yields (in the chiral limit of $m_\pi=0$) the result~\cite{BW-ERA-LC2019,Courtoy:2019cxq,Broniowski:2019rmu}
\begin{eqnarray}
D_{u\bar{d}}(x_1,x_2,\vect{q})= 1 \times \delta(1-x_1-x_2) \Theta   F(\vec{q}), \label{eq:dp}
\end{eqnarray}
where the $\delta$ function reflects the conservation of the  ${}^+$ components of the momentum and the $\Theta$ indicates the proper support 
$0 \le x_1,x_2 \le 1$. The form factor $ F(\vec{q})$ depends on the adopted regularization scheme, which 
is necessary to remove the hard momentum contribution from the model. Analogously to the case of sPDF, the distribution in $x_1$ or $x_2$ is uniform. 
The factorization of the longitudinal and transverse dynamics holds in the strict chiral limit.  

\begin{figure}
\begin{center}
\resizebox{0.49\columnwidth}{!}{\includegraphics{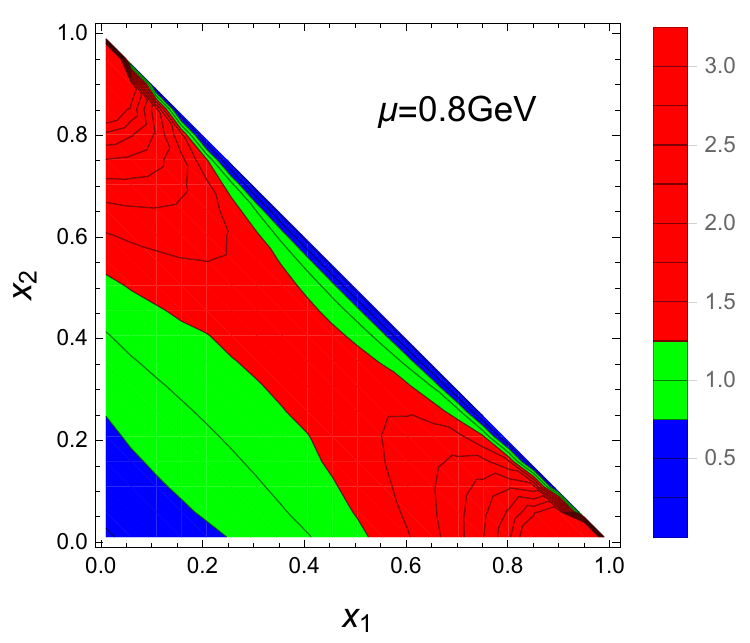}} \resizebox{0.49\columnwidth}{!}{\includegraphics{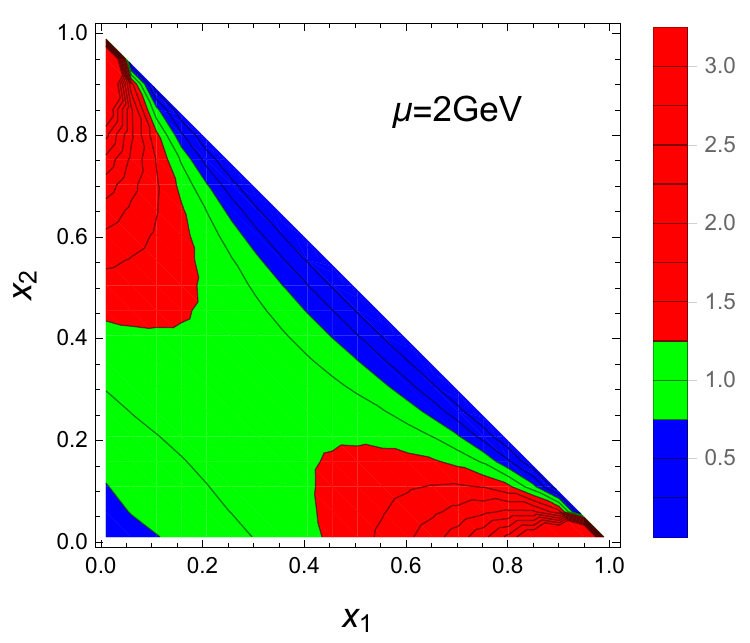}} \\
\resizebox{0.49\columnwidth}{!}{\includegraphics{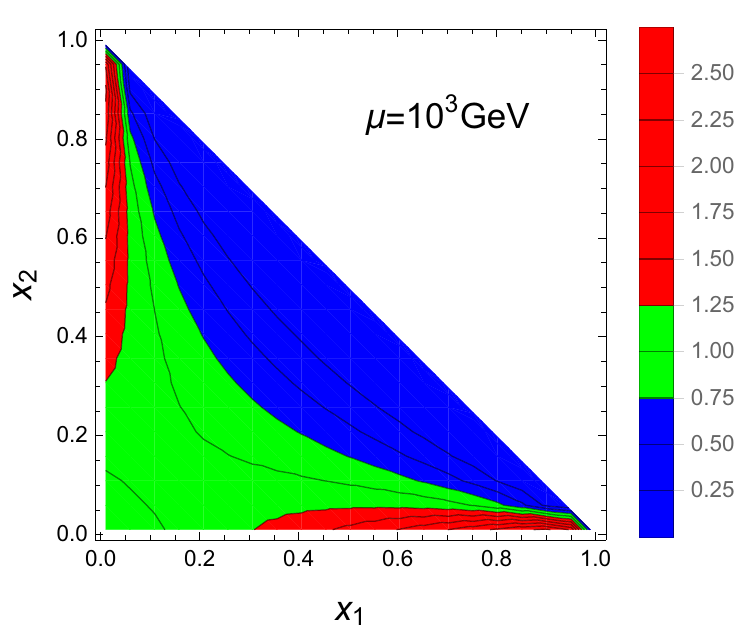}} \resizebox{0.49\columnwidth}{!}{\includegraphics{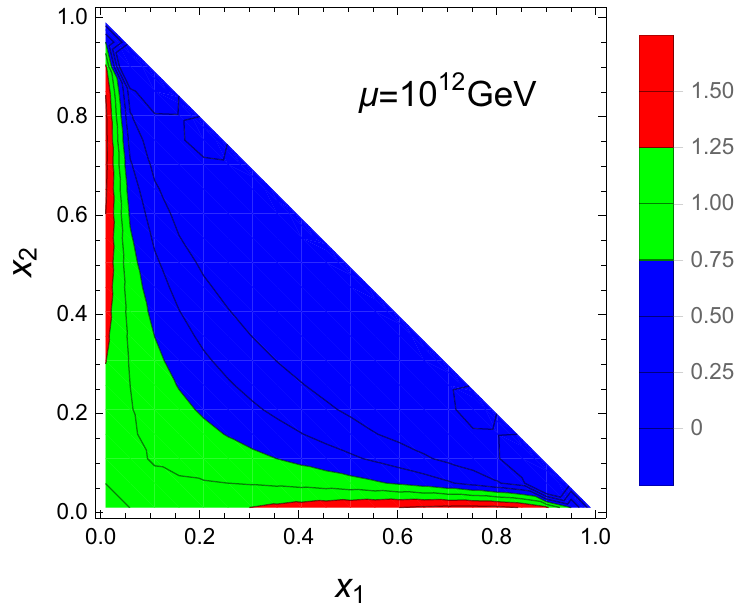}}
\end{center}
\caption{Correlation $D_{\rm u \bar{d}}(x_1,x_2)/D_u(x_1)D_{\bar{d}}(x_2)$ at various evolution scales $\mu$ indicated in the panels. 
The regions marked in green have the correlation within 20\% of unity.
(taken from~\cite{Broniowski:2019rmu}). \label{fig:corr}}
\end{figure}

First, let us bring up a very general formal aspect of dPDFs, namely, the Gaunt-Stirling (GS) sum rules~\cite{Gaunt:2009re}. 
These identities hold for the special case of $\vec{q}=0$, 
\begin{eqnarray}
 D_{ij}(x_1,x_2, \vec{q}=0) \equiv D_{ij}(x_1,x_2), \label{eq:not}
\end{eqnarray}
and link the marginal projections/moments of dPDFs to sPDFs in a way
typical of probability distributions. They follow from a decomposition of the
parton operators in a basis of the light-front wave functions~\cite{Gaunt:2012tfk}
\begin{eqnarray}
&&\sum_{i} \int_0^{1-x_2} \!dx_1 \,x_1 D_{ij}(x_1,x_2)=(1-x_2) D_{j}(x_2), \nonumber \\
&&\int_0^{1-x_2} \! dx_1 \, D_{i_{\rm val} j}(x_1,x_2) = (N_{i_{\rm val}}-\delta_{ij}+\delta_{\bar{i}j})D_j(x_2), \label{eq:gs}
\end{eqnarray}
where $i_{\rm val}$ is the difference of the parton ($i$) and anti-parton ($\bar{i}$) distributions, with $N_{i_{\rm val}}=\int_0^1 dx \, D_{i_{\rm val}}(x)$.
In numerous approaches on the market, the GS sum rules are non-trivial to satisfy~\cite{Gaunt:2009re,Golec-Biernat:2014bva,Diehl:2018kgr,Diehl:2020xyg}.
We have shown in~\cite{Broniowski:2013xba,Broniowski:2016trx} that a successful approach follows from a top-down method,
starting from  an  $n$-parton distributions with \mbox{$\delta(1-x_1-...-x_n)$} from momentum conservation, 
which yields the distributions with lower number of partons via subsequent marginal projections. 

We note that the NJL form of Eq.~(\ref{eq:dp}) explicitly satisfies the GS sum rules, as the model obeys all constraints from the Lorentz covariance and symmetries.
Furthermore, as the sum rules hold when the QCD evolution is applied (see below), the feature holds at any scale. 

The scheme to evolve dPDFs has been
derived long ago~\cite{Kirschner:1979im,Shelest:1982dg}. An efficient method is based on the Mellin moments, similarly
to the case of sPDFs. Details are presented in~\cite{Broniowski:2013xba,Broniowski:2019rmu}.

Our results for the valence dPDF of the pion (conventionally multiplied with $x_1 x_2$), 
evolved to increasing scales $\mu$,  are shown in Fig.~\ref{fig:dev}.
We note  that, as expected from the DGLAP evolution, the strength shifts to lower $x_1$ and $x_2$ as $\mu$ increases. 
Note that the initial condition for the evolution is distributed singularly along the $x_1=x_2$ line, as given in Eq.~(\ref{eq:dp}). 
The evolution washes out this behavior, filling all the triangle with $x_1+x_2 \le 1$. Of course, as the evolution generates radiatively more partons, 
the two valence partons do not need to carry all the momentum, hence $x_1+x_2$ can be smaller than~1.

Next, in Fig.~\ref{fig:corr} we show our results for the correlation $D_{\rm u \bar{d}}(x_1,x_2)/D_u(x_1)D_{\bar{d}}(x_2)$. 
This is an important measure, as whenever it departs significantly from~1, one cannot use the factorization assumption frequently made in 
analyses of the double parton scattering (see~\cite{Bartalini:2017jkk} for a recent review). The regions in the plots marked with green have the 
correlation within 25\% from the unity, indicated the region where factorization approximately works, whereas outside of this region it is significantly broken. 
We note that at low $x_1$ and $x_2$ increasing the evolution scale brings the correlation ratio closer to~1, which is in line with the 
conclusions of~\cite{Golec-Biernat:2015aza} for the gluon distributions in the nucleon. 

\begin{figure}
\begin{center}
\resizebox{0.52\columnwidth}{!}{\includegraphics{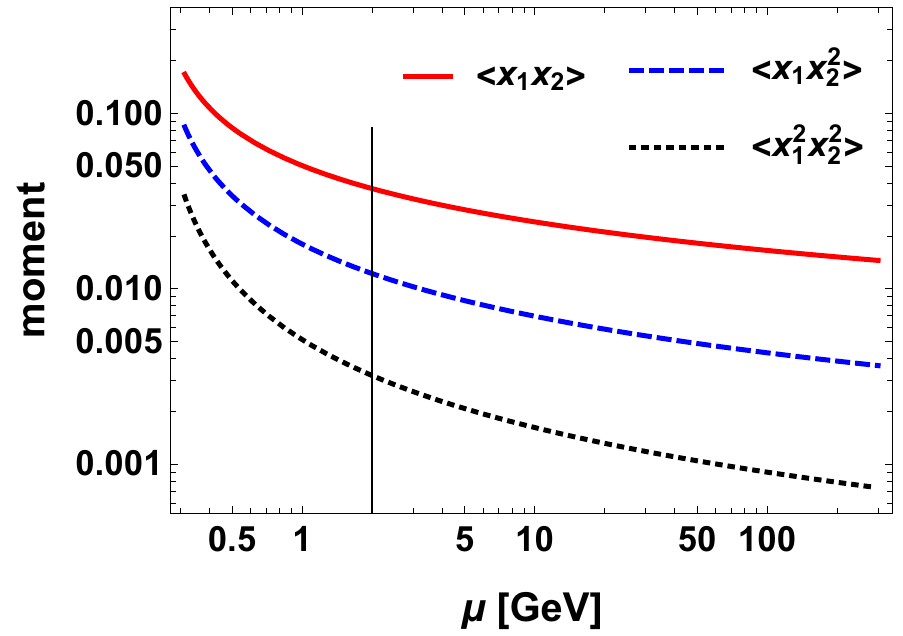}}
\end{center}
\caption{Lowest Mellin moments of the valence dPDF of the pion, plotted as functions of the evolution scale. The vertical line indicates the typical lattice scale of 2~GeV.
\label{fig:mom-evol}} 
\end{figure}

As proposed in~\cite{Broniowski:2019rmu}, a very practical measures of correlation 
which hopefully could be probed in the upcoming lattice analyses, 
are based on the ratios of the Mellin moments for the valence 
distributions,
\begin{eqnarray}
\frac{\langle  x_1^n x_2^m \rangle}{\langle  x_1^n \rangle \langle x_2^m \rangle}. \label{eq:momdef}
\end{eqnarray}
The key point is that these ratios are independent of the evolution scale, since the corresponding factors involving the anomalous dimensions cancel out. 
This feature occurs only for the valence case, where the evolution does not mix dDPFs and sPDFs via the so called inhomogeneous term. 
In the NJL model, we have the simple formula 
\begin{eqnarray}
\frac{\langle  x_1^n x_2^m \rangle}{\langle  x_1^n \rangle \langle x_2^m \rangle}= \frac{(1+n)!(1+m)!}{(1+n+m)!}. \label{eq:momnjl}
\end{eqnarray}
The lowest values of these ratios are listed in Table~\ref{tab:rat}, and the evolution of a few lowest moments is presented in Fig.~\ref{fig:mom-evol}.

\begin{table}
\label{tab:rat}
\caption{The ratios of the double to single valence moments,  $\langle  x_1^n x_2^m \rangle /\langle  x_1^n \rangle \langle x_2^m \rangle$, in the NJL 
model, Eq.~(\ref{eq:momnjl}).
Rows and columns correspond to $n$ and $m$. These ratios are independent of the evolution scale $\mu$.}
\begin{center}
\begin{tabular}{c|cccc}
   & 1 & 2 & 3 & 4  \\ \hline
 1 & $ {2}/{3}$ & $ {1}/{2}$ & $ {2}/{5}$ & $ {1}/{3}$  \\
 2 & $ {1}/{2}$ & $ {3}/{10}$ & $ {1}/{5}$ & $ {1}/{7}$  \\
 3 & $ {2}/{5}$ & $ {1}/{5}$ & $ {4}/{35}$ & $ {1}/{14}$  \\
 4 & $ {1}/{3}$ & $ {1}/{7}$ & $ {1}/{14}$ & ${5}/{126}$ 
\end{tabular}
\end{center}
\end{table}

\begin{figure}
\begin{center}
\resizebox{0.52\columnwidth}{!}{\includegraphics{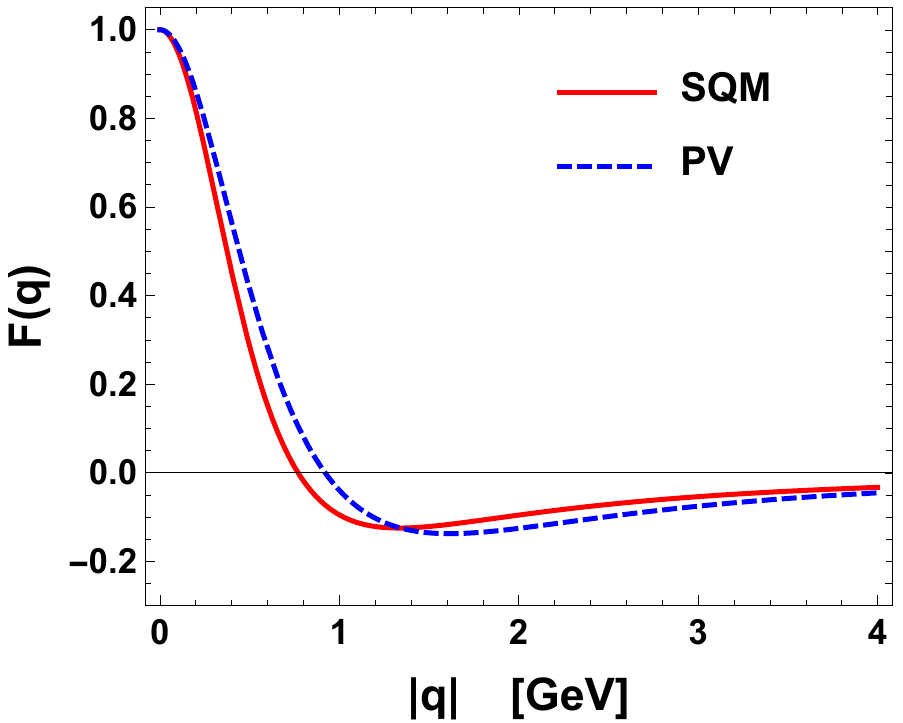}}
\end{center}
\caption{Valence dPDF form factor of the pion in the SQM (solid line) and in NJL with PV regularization (dashed line), plotted
as a function of the transverse momentum.
\label{fig:ff}} 
\end{figure}

Finally, we discuss the transverse form factor $F(\vec{q})$, appearing in Eq.~(\ref{eq:dp}). 
Its form depends on the regularization scheme of the NJL model, and for fundamental reasons can only be trusted for soft external momenta. 
In the Spectral Quark Model (SQM)~\cite{RuizArriola:2003bs} it has a very simple form
\begin{eqnarray}
F(\vect{q}) = \frac{m_\rho^4-\vect{q}^2 m_\rho^2}{\left(m_\rho^2+\vect{q}^2\right){}^2}. \label{eq:FF}
\end{eqnarray}
In the NJL model with the Pauli-Villars (PV) subtraction a numerically
very similar result follows, as can be seen from Fig.~\ref{fig:ff}.
The region with $|\vect{q}|>1$~GeV, where the form factor goes negative, is
outside of the validity of the model.

The corresponding effective cross section for the double parton scattering coincides (in SQM) with the geometric cross section, namely
\begin{eqnarray}
\sigma_{\rm eff} = \frac{1}{\int \frac{d^2\vect{q}_\perp}{(2 \pi)^2} 
F (\vect{q}_\perp) F (-\vect{q}_\perp) } =  \pi \frac{12}{m_\rho^2} = \pi \langle b^2 \rangle = 23~{\rm mb}. \label{eq:seff}
\end{eqnarray}

\section{Summary}
\label{sec:sum}

To summarize, here are our main points:

\begin{itemize}
   
\item The topic of the double parton correlations is driven by recent experimental evidence as well as by possible future lattice studies.
   
\item Our results are obtained in the NJL model which is a simple field theory of the pion in the soft regime based on the spontaneous chiral symmetry breaking.
 It is a covariant calculation, with all symmetries preserved, which leads to proper formal features. In particular, the GS sum rules are satisfied. The QCD 
 evolution to higher scales is a crucial ingredient of the approach.
 
 \item The model leads, in the chiral limit, to the longitudinal-transverse 
    factorization, whereas there is no factorization in the $x_1$ and $x_2$ variables, which are correlated due to the momentum conservation. 
    
 \item The correlation ratio at low $x_1$ and $x_2$ is brought close to 1 with increasing evolution scale.
     
 \item The appropriate ratios of the  Mellin moments do not depend on the evolution scale, hence are particularly convenient. 
 They could be probed in future lattice simulations.

\end{itemize}

\bigskip

Both authors have contributed equally to all stages of the research and preparation of this article.

\bibliographystyle{JHEP}
\bibliography{dPDF,refs,covid-19}

\end{document}